\input harvmac
\input epsf
\noblackbox
\newcount\figno
\figno=0
\def\fig#1#2#3{
\par\begingroup\parindent=0pt\leftskip=1cm\rightskip=1cm\parindent=0pt
\baselineskip=11pt
\global\advance\figno by 1
\midinsert
\epsfxsize=#3
\centerline{\epsfbox{#2}}
\vskip 12pt
\centerline{{\bf Figure \the\figno} #1}\par
\endinsert\endgroup\par}
\def\figlabel#1{\xdef#1{\the\figno}}
\def\pano{\par\noindent}

\font\cmss=cmss10
\font\cmsss=cmss10 at 7pt

\def\rlx{\relax\leavevmode}
\def\inbar{\vrule height1.5ex width.4pt depth0pt}
\def\IC{\relax\,\hbox{$\inbar\kern-.3em{\rm C}$}}
\def\IR{\relax{\rm I\kern-.18em R}}
\def\IN{\relax{\rm I\kern-.18em N}}
\def\IP{\relax{\rm I\kern-.18em P}}
\def\ZZ{\rlx\leavevmode\ifmmode\mathchoice{\hbox{\cmss Z\kern-.4em Z}}
 {\hbox{\cmss Z\kern-.4em Z}}{\lower.9pt\hbox{\cmsss Z\kern-.36em Z}}
 {\lower1.2pt\hbox{\cmsss Z\kern-.36em Z}}\else{\cmss Z\kern-.4em Z}\fi}
\def\narrowplus{\kern -.04truein + \kern -.03truein}
\def\narrowminus{- \kern -.04truein}
\def\narrowminussub{\kern -.02truein - \kern -.01truein}

\def\o#1{\overline{#1}}

\def\th#1#2{\vartheta\bigl[{\textstyle{  #1 \atop #2}} \bigr] }

\def\kl#1{\left(#1\right)}



\lref\rseiwit{N. Seiberg and  E. Witten, {\it String Theory and 
Noncommutative Geometry}, JHEP {\bf 9909} (1999) 032, hep-th/9908142.}

\lref\rschomi{V. Schomerus, {\it D-branes and Deformation Quantization},
JHEP {\bf 9906} (1999) 030, hep-th/9903205.}

\lref\rangles{M. Berkooz, M.R. Douglas and R.G. Leigh, {\it Branes Intersecting 
at Angles}, Nucl. Phys. {\bf B480} (1996) 265, hep-th/9606139.}

\lref\noncom{
C.G. Callan, C. Lovelace, C.R. Nappi and S.A. Yost, {\it
String Loop Corrections To Beta Functions}, Nucl. Phys. {\bf B288}
(1987) 525;
A. Abouelsaood, C.G. Callan, C.R. Nappi and S.A. Yost, 
{\it Open Strings in Background Gauge Fields}, Nucl. Phys. {\bf B280}
(1987) 599;
Y.-K.E. Cheung and M. Krogh, {\it Noncommutative Geometry from 
D0-branes in a Background B-field}, Nucl.Phys. {\bf B528} (1998) 185,
hep-th/9803031;
J. Fr\"ohlich, O. Grandjean and A. Recknagel, 
{\it Supersymmetric Quantum Theory and Noncommutative Geometry}, 
Commun. Math. Phys. {\bf 203} (1999) 119, math-ph/9807006;
F. Ardalan, H. Arfaei and  M. M. Sheikh-Jabbari, 
{\it
Dirac Quantization of Open Strings and Noncommutativity in Branes},
hep-th/9906161;
C.-S. Chu and  P.-M. Ho, 
{\it Noncommutative Open String and D-brane},  
Nucl.Phys. {\bf B550} (1999) 151, hep-th/9812219;
{\it Constrained Quantization of Open 
String in Background B Field and Noncommutative D-brane}, hep-th/9906192;
J. Madore, S. Schraml, P. Schupp and J. Wess,
{\it Gauge theory on noncommutative spaces},
hep-th/0001203;
B. Jurco, P. Schupp and J. Wess,
{\it Noncommutative gauge theory for Poisson manifolds}, hep-th/0005005.
}

\lref\jab{F. Ardalan, H. Arfaei and  M. M. Sheikh-Jabbari, {\it 
Noncommutative Geometry From Strings and Branes}, 
JHEP {\bf 9902} (1999) 016, hep-th/9810072.}

\lref\rconnes{A. Connes, M.R. Douglas and  A. Schwarz, {\it 
Noncommutative Geometry and Matrix Theory: Compactification on Tori},
JHEP {\bf 9802} (1998) 003, hep-th/9711162;
M.R. Douglas and  C. Hull, {\it
D-branes and the Noncommutative Torus}, JHEP {\bf 9802} (1998) 008,
hep-th/9711165.}

\lref\wittena{E. Witten, {\it Search for realistic Kaluza-Klein theory},
Nucl. Phys. {\bf B186} (1981) 412.}

\lref\CHSW{P. Candelas, G. Horowitz, A. Strominger and E. Witten,
{\it Vacuum configurations for superstrings}, Nucl. Phys. {\bf B258} (1985) 46.}

\lref\LAT{W. Lerche, D. L\"ust and A.N. Schellekens, {\it Chiral, 
four-dimensional heterotic strings from self-dual lattices},
Nucl. Phys. {\bf B287} (1987) 477.}

\lref\FERM{H. Kawai, D. Lewellen and S.H. Tye, {\it Construction of fermionic
string models in four dimensions}, Nucl. Phys. {\bf B288} (1987) 1;
I. Antoniadis, C. Bachas and C. Kounnas, {\it Four-dimensional superstrings},
Nucl. Phys. {\bf B289} (1987) 87.}

\lref\SING{M. Douglas and G. Moore, {\it D-branes, quivers and ALE instantons},
hep-th 9603167;
I.R. Klebanov and E. Witten, {\it Superconformal field theory on three-branes
at a Calabi-Yau singularity}, Nucl. Phys. {\bf B536} (1998) 199,
hep-th/9807080.}

\lref\NS{K. Landsteiner, E. Lopez and D. Lowe, {\it Duality of chiral $N=1$
supersymmetric gauge theories via branes}, JHEP {\bf 9802} (1998) 007,
hep-th/9801002;
I. Brunner, A. Hanany, A. Karch and D. L\"ust,
{\it Brane dynamics and chiral non-chiral transitions},
Nucl. Phys. {\bf B528} (1998) 197, hep-th/9801017;
A. Hanany and A. Zaffaroni, {\it On the realization of chiral four-dimensional
gauge theories using branes}, JHEP {\bf 9805} (1998) 001, 
hep-th/9801134.}

\lref\HORWIT{P. Horava and E. Witten, {\it Heterotic and type I string dynamics
from eleven dimensions}, Nucl. Phys. {\bf B460} (1996) 506, 
hep-th/9510209.}

\lref\ORBI{L. Dixon, J.A. Harvey, C. Vafa and E. Witten, {\it Strings on
orbifolds}, Nucl. Phys. {\bf B261} (1985) 678.}

\lref\FAUX{P. Horava and E. Witten, {\it Heterotic and type I string dynamics
from eleven dimensions}, Nucl. Phys. {\bf B460} (1996) 506, 
hep-th/9510209;
E. Witten, {\it Strong coupling expansion of  Calabi-Yau compactification},
Nucl. Phys. {\bf B471} (1996) 135, hep-th/9602070;
M. Faux, D. L\"ust and B.A. Ovrut, {\it Intersecting orbifold
planes and local anomaly cancellation in M-theory},
Nucl. Phys. {\bf B554} (1999) 437, hep-th/9903028;
V. Kaplunovsky, J. Sonnenschein, S. Theisen and S. Yankielowicz,
{\it On the duality between perturbative orbifolds and M-theory
on $T^4/Z_N$}, hep-th/9912144;
M. Faux, D. L\"ust and B.A. Ovrut, {\it Local Anomaly Cancellation,
M-theory orbifolds and phase transitions}, hep-th/0005251.}

\lref\GAUGINO{J.P. Derendinger, L.E. Ibanez and H.P. Nilles,
{\it On the low-energy $D=4$, $N=1$ supergravity theory extracted from the
$D=10$, $N=1$ superstring}, Phys. Lett. {\bf B155} (1985) 65;
M. Dine, R. Rohm, N. Seiberg and E. Witten,
{\it Gluino condensation in superstring models},
Phys. Lett. {\bf B156} (1985) 55.}

\lref\BB{I. Antoniadis, E. Dudas and A. Sagnotti, {\it Brane
supersymmetry breaking}, Phys. Lett. {\bf B464} (1999) 38, 
hep-th/9908023;
G. Aldazabal and A.M. Uranga, {\it Tachyon-free
Non-supersymmetric Type IIB Orientifolds via Brane-Antibrane Systems}, 
JHEP {\bf 9910} (1999) 024, hep-th/9908072;
G. Aldazabal, L.E. Ibanez and  F. Quevedo, {\it Standard-like
Models with Broken Supersymmetry from Type I String Vacua}, JHEP 
{\bf 0001} (2000) 031, hep-th/9909172; {\it A D-Brane
Alternative to the MSSM }, hep-ph/0001083;
C. Angelantonj, I. Antoniadis, G. D'Appollonio, E. Dudas,
A. Sagnotti, {\it Type I vacua with brane supersymmetry breaking},
hep-th/9911081;
C. Angelantonj, R. Blumenhagen and M.R. Gaberdiel,
{\it Asymmetric orientifolds, brane supersymmetry breaking and non BPS branes},
hep-th/0006033.}

\lref\VECTOR{E. Witten; {\it New issues in manifolds of $SU(3)$ holonomy},
Nucl. Phys. {\bf B268} (1986) 79.}

\lref\WILSON{L.E. Ibanez, H.P. Nilles and F. Quevedo,
{\it Reducing the rank of the gauge group in orbifold
compactifications of the heterotic string}, Phys. Lett. {\bf B192} (1987) 332.}

\lref\SM{I. Antoniadis, E. Kiritsis and T.N. Tomaras,
{\it A D-brane alternative to unification},  hep-th/0004214;
G. Aldazabal, L. E. Ibanez, F. Quevedo and A. M. Uranga, 
{\it D-Branes at Singularities : A Bottom-Up Approach to the String
Embedding of the Standard Model}, hep-th/0005067;
A. Krause, {\it A small cosmological constant, grand unification and warped
geometry}, hep-th/0006226.}

\lref\rbgkl{R. Blumenhagen, L. G\"orlich, B. K\"ors and D. L\"ust, 
{\it Asymmetric Orbifolds, Noncommutative Geometry and Type I
Vacua}, hep-th/0003024.}

\lref\rbgka{R. Blumenhagen, L. G\"orlich and B. K\"ors, 
{\it Supersymmetric Orientifolds in 6D with D-Branes at Angles},
Nucl. Phys. {\bf B569} (2000) 209, hep-th/9908130; 
{\it A New Class of Supersymmetric Orientifolds with D-Branes at
Angles}, hep-th/0002146;
{\it
Supersymmetric 4D Orientifolds of Type IIA with D6-branes at Angles},  
JHEP {\bf 0001} (2000) 040, hep-th/9912204.}

\lref\rba{C. Angelantonj, R. Blumenhagen, {\it Discrete Deformations in 
Type I Vacua}, Phys. Lett. {\bf B473} (2000) 86, 
hep-th/9911190.} 

\lref\rchu{C.-S. Chu, {\it Noncommutative Open String: Neutral and Charged},
hep-th/0001144.}

\lref\rchen{B. Chen, H. Itoyama, T. Matsuo and K. Murakami, {\it
p-p' System with B-field, Branes at Angles and Noncommutative Geometry}, 
{\tt hep-th/9910263}.}

\lref\raadds{C. Angelantonj, I. Antoniadis, G. D'Appollonio, E. Dudas,
A. Sagnotti, {\it Type I vacua with brane supersymmetry breaking},
hep-th/9911081, to appear in Nucl. Phys. {\bf B}.}

\lref\rads{I. Antoniadis, E. Dudas, A. Sagnotti, {\it Brane
supersymmetry breaking}, Phys. Lett. {\bf B464} (1999) 38, 
hep-th/9908023.}

\lref\rbfl{C. Angelantonj, {\it Non-tachyonic open descendants of the
0B string theory}, Phys. Lett. {\bf B444} (1998) 309, 
hep-th/9810214; R. Blumenhagen, A. Font and D. L\"ust, {\it Tachyon free
orientifolds of type 0B strings in various dimensions},
Nucl. Phys. {\bf B558} (1999) 159, hep-th/9904069;
R. Blumenhagen, A. Font, A. Kumar and D. L\"ust, {\it Aspects of type 0
string theory}, Class. Quant. Grav. {\bf 17} (2000) 989, hep-th/9908155;
R. Blumenhagen and A. Kumar, {\it A Note on Orientifolds and Dualities of Type
0B String Theory}, Phys.Lett. {\bf B464} (1999) 46, hep-th/9906234.}

\lref\rss{J. Scherk, J.H. Schwarz, {\it Spontaneous breaking of
supersymmetry through dimensional reduction}, Phys. Lett. {\bf B82}
(1979) 60; {\it How to get masses from extra dimensions}, 
Nucl. Phys. {\bf B153} (1979) 61;
E. Cremmer, J. Scherk, J.H. Schwarz, {\it Spontaneously
broken ${\cal N}=8$ supergravity}, Phys. Lett. {\bf B84} (1979) 83;
S. Ferrara, C. Kounnas, M. Porrati, F. Zwirner, {\it Superstrings
with spontaneously broken supersymmetry and their effective theories},
Nucl. Phys. {\bf B318} (1989) 75.} 

\lref\rssst{R. Rohm, {\it Spontaneous supersymmetry breaking in
supersymmetric string theories}, Nucl. Phys. {\bf B237} (1984) 553; 
C. Kounnas, M. Porrati, {\it Spontaneous supersymmetry breaking in
string theory}, Nucl. Phys. {\bf B310} (1988) 355; 
S. Ferrara, C. Kounnas, M. Porrati, F. Zwirner, {\it Superstrings
with spontaneously broken supersymmetry and their effective theories},
Nucl. Phys. {\bf B318} (1989) 75; 
C. Kounnas, B. Rostand, {\it Coordinate dependent compactifications
and discrete symmetries}, Nucl. Phys. {\bf B341} (1990) 641;
I. Antoniadis, C. Kounnas, {\it Superstring phase transition at
high temperature}, Phys. Lett. {\bf B261} (1991) 369;
E. Kiritsis, C. Kounnas, {\it Perturbative and non-perturbative
partial supersymmetry breaking: ${\cal N} = 4 \to {\cal N} = 2 \to
{\cal N} =1$}, Nucl. Phys. {\bf B503} (1997) 117, 
hep-th/9703059.} 

\lref\wise{J. Gomis, T. Mehen and M.B. Wise,
{\it Quantum field theories with compact noncommutative extra dimensions},
hep-th/0006160.}

\lref\rssop{I. Antoniadis, E. Dudas, A. Sagnotti, 
{\it Supersymmetry breaking, open strings and M-theory},
Nucl. Phys. {\bf B544} (1999) 469, hep-th/9807011; 
I. Antoniadis, G. D'Appollo\-nio, E. Dudas, A. Sagnotti, {\it Partial
breaking of supersymmetry, open strings and M-theory},
Nucl. Phys. {\bf B553} (1999) 133, hep-th/9812118; 
{\it Open descendants of $\ZZ_2 \times \ZZ_2$ freely acting
orbifolds}, Nucl. Phys. {\bf B565} (2000) 123, hep-th/9907184.} 

\lref\rmb{A. Sagnotti, {\it Anomaly cancellations and open-string
theories}, hep-th/9302099; 
G. Zwart, {\it Four-dimensional ${\cal N}=1$ $\ZZ_N \times \ZZ_M$
orientifolds}, Nucl. Phys. {\bf B526} (1998) 378, 
hep-th/9708040; 
Z. Kakushadze, G. Shiu, S.-H.H. Tye, {\it Type IIB orientifolds,
F-theory, type I strings on orbifolds and type I - heterotic duality},
Nucl. Phys. {\bf B533} (1998) 25, {\tt hep-th/9804092};
G. Aldazabal, A. Font, L. E. Ibanez, G. Violero, {\it $D=4$, ${\cal
N} =1$ type IIB orientifolds}, Nucl. Phys. {\bf B536} (1998) 29,
{\tt hep-th/9804026}.}

\lref\rau{G. Aldazabal, A.M. Uranga, {\it Tachyon-free
non-supersymmetric type IIB orientifolds via brane-antibrane systems}, 
JHEP {\bf 9910} (1999) 024, hep-th/9908072.}

\lref\rbac{C. Bachas, {\it A way to break supersymmetry},
hep-th/9503030.}
 
\lref\rbist{M. Bianchi, Ya.S. Stanev, {\it Open strings on the Neveu-Schwarz
penta-brane}, Nucl. Phys. {\bf B523} (1998) 193, hep-th/9711069.}

\lref\abpss{C. Angelantonj, M. Bianchi, G. Pradisi, A. Sagnotti and
Y. Stanev, {\it Chiral asymmetry in four-dimensional open string vacua},
Phys. Lett. {\bf B385} (1996) 96, hep-th/9606169.}

\lref\AAD{C. Angelantonj, I. Antoniadis, E. Dudas and A. Sagnotti, 
in preparation.}

\Title{\vbox{
 \hbox{HU--EP--00/27}
 \hbox{hep-th/0007024}}}
{\vbox{\centerline{Noncommutative Compactifications of Type I Strings}
       \bigskip
       \centerline{on Tori with Magnetic Background Flux} }}
\centerline{Ralph Blumenhagen\footnote{$^1$}{{\tt e-mail:
blumenha@physik.hu-berlin.de}}, Lars G\"orlich\footnote{$^2$}{{\tt e-mail:
goerlich@physik.hu-berlin.de}}, Boris K\"ors\footnote{$^3$}{{\tt e-mail:
koers@physik.hu-berlin.de}} and Dieter L\"ust\footnote{$^4$}{{\tt e-mail:
luest@physik.hu-berlin.de}}} 
\bigskip
\centerline{\it Humboldt-Universit\"at zu Berlin, Institut f\"ur  
Physik,}
\centerline{\it Invalidenstrasse 110, 10115 Berlin, Germany}
\smallskip
\bigskip
\centerline{\bf Abstract}
\noindent
We construct six- and four-dimensional  
toroidal compactifications of the Type I string with magnetic flux on the 
D-branes. The open strings in this  background probe a
noncommutative internal geometry. Phenomenologically appealing features 
such as chiral fermions and supersymmetry breaking in the gauge sector
are naturally realized by these vacua. 
We investigate the spectra of such noncommutative string compactifications
and in a bottom-up approach discuss the possibility to obtain the standard
or some GUT like model.
\bigskip

\Date{06/2000}

\newsec{Introduction}

The search for realistic string vacua is one of the burning open problems
within superstring theory.
A phenomenologically viable string compactification should contain at least
three chiral fermion generations, the standard model gauge group and
broken space-time supersymmetry. In the context of `conventional'
string compactifications the requirement of getting chiral fermions
is usually achieved by considering compact, internal background spaces
with nontrivial topology rather than simple tori.
In particular, when analyzing the Kaluza-Klein fermion spectra \wittena\
a net-fermion generation number arises if the internal Dirac operator
has zero modes. For example, considering heterotic string compactifications
on Calabi-Yau threefolds \CHSW, 
the net-generation number is equal to $|\chi|/2$,
where $\chi$ is the Euler number of the Calabi-Yau space. 
Chiral fermions are also present in a large
class of heterotic orbifold compactifications \ORBI, 
as well as in free bosonic \LAT\
and fermionic \FERM\ constructions. 
Type II string models with chiral fermions can be constructed by locating
D-branes at transversal orbifold or conifold singularities \SING,
or by considering intersections of D-branes and NS-branes \NS;
chiral type I models were first proposed in \abpss.
Moreover 
orbifold compactificactions of eleven-dimensional M-theory
can lead to chiral fermions, as discussed e.g. in
\FAUX.

The phenomenological requirement of breaking space-time supersymmetry can
be met in various ways. In the context of heterotic string compactifications
gaugino condensation \GAUGINO\ or the Scherk-Schwarz mechanism 
\refs{\rss,\rssop}
lead to potentially interesting models with supersymmetry broken at low
energies. In addition, as it was realized more recently, type II models
on nontrivial background spaces with certain D-brane configurations 
possess broken space-time supersymmetry. Especially, when changing the
GSO-projections tachyon free type 0 orientifolds 
in four dimensions can be constructed \rbfl. 
Alternatively, orientifolds on six-dimensional orbifolds with brane-antibrane
configurations provide interesting scenarios \BB, where supersymmetry is
left unbroken in the gravity bulk, but broken in the open string sector
living on the brane-antibrane system.

Finally the quest for a realistic gauge group with sufficiently low rank
is met in heterotic strings by choosing appropriate gauge vector bundles
on the Calabi-Yau spaces \VECTOR, 
which can be alternatively described by turning on Wilson lines
in Calabi-Yau or also in orbifold compactifications \WILSON.
On the type II side the rank of the gauge group can be also lowered 
by Wilson lines or, in the T-dual picture, by placing the branes at
different positions inside the internal space.

As it should have become clear from the previous discussion, `standard'
heterotic, type I or type 
II compactifications on simple 6-tori do not meet any of the three above
requirements. However, as we will discuss in this paper, turning on 
magnetic fluxes in the internal directions of the D-branes, thereby inducing 
mixed Neumann-Dirichlet boundary conditions for open strings equivalent to a 
noncommutative internal geometry \rconnes\
on the branes, all three goals can be
achieved in one single stroke.\footnote{$^1$}{In a previous paper \rbgkl\
we have discussed type I string compactifications on noncommutative
asymmetric orbifold spaces.} 
Specifically, we will discuss type I string compactifications on a product
of $d$ noncommutative two-tori to $10-2d$ non-compact Minkowski dimensions
($d=2,3$),
i.e. the ten-dimensional
background spaces $M^{10}$ we are considering have the following form
\eqn\backgr{M^{10}=R^{1,9-2d}\times\prod_{j=1}^d T^2_{(j)}.}
The coordinates in the internal space possess the following commutation
relations 
\eqn\commutator{ [ X_{10-2j} ,X_{11-2j} ]=i\theta^{(j)}\, ,
\qquad j=1,\dots , d.
}
These commutation relations will be realized by D9-branes with constant
background
magnetic fluxes $F^{(j)}$ turned on in the directions of the 2-tori
\refs{\noncom\rschomi\rseiwit-\jab}, corresponding to the following 
noncommutative deformation parameter $\theta^{(j)}$ in eq.\commutator:
\eqn\defpar{
\theta^{(j)}=-{2\pi  \alpha'  F^{(j)} \over 
      1+(F^{(j)})^2}.} 
The entire internal noncommutative torus will actually consist out of different
sectors with different noncommutative deformation parameters, because
we will introduce several D9-branes  with different magnetic fluxes.
We will show that the spectrum of open strings,
with mixed boundary
conditions in the
internal directions is generically chiral, breaks space-time supersymmetry and
leads to gauge groups of lower rank.
It is however important to stress
that
the effective gauge theories in the uncompactified part of space-time
are still commutative, and therefore are Lorentz invariant and local
field theories.


This construction is the D-brane extended version of 
\rbac, where
it was already observed that turning on magnetic flux in a toroidal
type I compactification leads to supersymmetry breaking and 
chiral massless spectra in four space-time dimensions. 
However, the consistency conditions for such models were derived
in the effective non-supersymmetric gauge theories, leaving the
actual string theoretic conditions an open issue. 
We will show that, with all the insights gained  
in the description of D-branes with magnetic flux, 
we are now able to achieve a complete string theoretic understanding, 
giving rise to certain extensions and modifications of the purely 
field theoretical analysis.  
As a solution to the tadpole cancellation conditions we can get
different sectors of D-branes with different magnetic fluxes, corresponding
to different noncommutative boundary conditions.  
Chirality then arises in sectors of open strings which have ends on branes 
with different gauge flux, while the presence of any solitary flux is 
not sufficient. The gauge groups that act on the D-branes with non-vanishing 
flux are unitary instead of orthogonal or symplectic in accord with the 
general statement that only these are compatible with a noncommutative 
deformation of the coordinate algebra.

For this kind of models, it is sometimes very helpful to employ an  
equivalent T-dual
description, where the background fields vanish and the 
torus is entirely commutative, but
the D-branes intersect at various different angles \rbgka. 
This description allows to
present  a more intuitive picture of the open string sector
involved in such models.  
Chiral fermions then arise
due to the nontrivial geometric boundary conditions of the intersecting 
D-branes, which at the same time also break space-time supersymmetry
and lower the rank of the gauge group. 

The paper is organized as follows. In the next section we analyze the
one-loop amplitudes and the resulting tadpole cancellation conditions
for D9-branes with mixed Neumann-Dirichlet boundary conditions 
moving in the background of $d$
two-dimensional tori ($d=2,3$). In section 3 we discuss specific
six-dimensional models ($d=2$) working out the non-supersymmetric,
chiral spectrum. We also point out some subtleties 
involving the mechanisms of supersymmetry breaking in `nearly' supersymmetric 
brane configurations. 
In section 4 we move on to chiral, non-supersymmetric
four-dimensional models ($d=3$), reconsider in particular the model 
presented in \rbac\ with GUT-like gauge group 
$G=U(5)\times U(3)\times U(4)
\times U(4)$ and display another 4 generation model with 
`standard model' gauge group
$G=U(3)\times U(2)\times U(1)^r$.\footnote{$^2$}{For other recent 
bottom up attempts
to obtain GUT's and the standard model from branes see \SM.}
Some phenomenological problems of this model are stressed at the end.

\newsec{One loop amplitudes}

In \rbac\ it was observed that turning on magnetic flux in a toroidal
type I compactification leads to supersymmetry breaking and in general
to chiral massless spectra in four space-time dimensions. 
The consistency conditions for such models were derived
in the effective non-supersymmetric gauge theory but not in the full 
string theory. 
In this section we will show that, with the inclusion of
D-branes with magnetic flux, respectively D-branes
at angles, we are now able to derive the string theoretic tadpole
cancellation conditions. 

\subsec{D9-branes with magnetic fluxes}

As our starting point we consider
the orientifold
\eqn\oria{   {{\rm Type\, IIB}\ {\rm  on}\ T^{2d} \over \Omega },  } 
where $\Omega$ denotes the word-sheet parity transformation. 
In the following we will assume that $T^{2d}$ splits into
a direct product of $d$ two-dimensional tori $T^2_{(j)}$ 
with coordinates $X^{(j)}_1$, $X^{(j)}_2$ and radii $R^{(j)}_1$, 
$R^{(j)}_2,\ j=1, \dots ,d$.  
Their complex structures will always be taken to be purely imaginary,
and 
the antisymmetric NSNS tensorfield will  be set to zero. 
Turning on magnetic flux, $F^{(j)}_{12}=F^{(j)}$, on 
a D9-brane changes the pure Neumann 
boundary conditions into mixed Neuman-Dirichlet conditions
\eqn\bounda{\eqalign{  
   &\partial_\sigma X^{(j)}_1 + F^{(j)} \partial_\tau X^{(j)}_2=0 ,\cr
   &\partial_\sigma X^{(j)}_2 - F^{(j)} \partial_\tau X^{(j)}_1=0 ,\cr}}
   
Let us consider different kinds of D9$_\mu$-branes, labelled by 
$\mu\in\{1,\ldots,K\}$,  distinguished by different magnetic fluxes 
on at least one torus. Specifically 
we are only considering branes which are characterized 
by $d$ sets of two  
integers $(n^{(j)}_\mu,m^{(j)}_\mu)$ corresponding to the 
electric respectively  magnetic charges\footnote{$^3$}
{All models considered in \rbac\ correspond to the subset of branes 
with $n^{(j)}_\mu=1$ for 
all $j$ and $\mu$.}. 

The magnetic fluxes  of such a brane  are given by
\eqn\angle{  F^{(j)}_\mu={m^{(j)}_\mu  \over 
n^{(j)}_\mu R^{(j)}_1R^{(j)}_2}.}
Following \rseiwit\ the boundary conditions eq.\bounda\ define an open string
metric on $T^2_{(j)}$ in the following way:
\eqn\openm{
G^{(j)kl}_\mu={1\over 1+(F^{(j)}_\mu)^2}\, \delta^{kl}.}
The deformation parameter of the noncommutative torus is given by
\eqn\deforpar{
\theta^{(j)kl}_\mu=-{2\pi\alpha' F^{(j)}_\mu\over 1+(F^{(j)}_\mu )^2}\,
\epsilon^{kl}.}

Let us now discuss the Kalazu-Klein mass spectrum of open strings 
on the noncommutative
torus. Since the translations on the torus are not deformed
by eq.\commutator, the center
of mass Kaluza-Klein momenta on $T^2_{(j)}$ are unchanged and are given by
\eqn\kkmom{
p^{(j)}_{\mu 1}={r^{(j)}_\mu\over  n^{(j)}_\mu R^{(j)}_1},
\qquad p^{(j)}_{\mu 2} ={s^{(j)}_\mu
\over n^{(j)}_\mu R^{(j)}_2}.}
Here the electric charge of the brane 
enters as the number of times it wraps the 
torus while the integers
$r^{(j)}_\mu$ and $s^{(j)}_\mu$ are the two Kaluza-Klein momentum numbers
along $T^2_{(j)}$.
Howover the masses of the open string Kalazu-Klein states do depend on the
magnetic fluxes.
Using the boundary state formalism one can compute the mass spectrum
of the open string Kalazu-Klein states 
on $T^2_{(j)}$ with mixed boundary conditions specified by the two integers
$n^{(j)}_\mu$ and $m^{(j)}_\mu$
with the following result \jab,\rbgkl:
\eqn\specb{  (M^{(j)}_\mu )^2={ (s^{(j)}_\mu R_1^{(j)})^2
+(r^{(j)}_\mu R_2^{(j)})^2\over  
                     (m^{(j)}_\mu )^2+ (n^{(j)}_\mu R_1^{(j)}R_2^{(j)})^2} .}
This allows use to extract the `noncommutative' Kaluza-Klein momenta and 
winding contributions
\eqn\momenta{\eqalign{ 
 \tilde p^{(j)}_\mu &
     ={n^{(j)}_\mu R_1^{(j)}R_2^{(j)}\over
    (m^{(j)}_\mu )^2+ (n^{(j)}_\mu R_1^{(j)}R_2^{(j)})^2}
    \left( r^{(j)}_\mu R_2^{(j)}+is^{(j)}_\mu R_1^{(j)} 
     \right) = 
    {1\over 1+(F^{(j)}_\mu)^2} 
    \left( p^{(j)}_{\mu 1} +i p^{(j)}_{\mu 2} \right) ,
  \cr
 w^{(j)}_\mu &=
   {m^{(j)}_\mu\over
    (m^{(j)}_\mu )^2+ (n^{(j)}_\mu R_1^{(j)}R_2^{(j)})^2}
    \left( -s^{(j)}_\mu R_1^{(j)}+ir^{(j)}_\mu R_2^{(j)} 
     \right) = 
   {F^{(j)}_\mu \over 1+(F^{(j)}_\mu)^2} 
    \left( -p^{(j)}_{\mu 2} +i p^{(j)}_{\mu 1} \right) , \cr }} 
(here written in complex notation)
where the mass formula eq.\specb\ is given as $(M^{(j)}_\mu )^2=
|\tilde p^{(j)}_\mu |^2+|w^{(j)}_\mu |^2$.
Of course, for $F^{(j)}_\mu =0$ the momenta $p^{(j)}_\mu$ and 
$\tilde p^{(j)}_\mu$ agree, and  $w^{(j)}_\mu =0$.
Note that the KK masses in eq.\specb\ can be also expressed by using the
standard KK momenta eq.\kkmom\ togother with the open string
metric eq.\openm\ as
\eqn\specc{
(M^{(j)}_\mu)^2= p^{(j)}_{\mu k}G^{(j)kl}_\mu p^{(j)}_{\mu l}.}

To convince ourselves that
we are indeed dealing with a noncommutative
torus for open strings  which end on the $D9_\mu$-brane
with magnetic flux $F^{(j)}_\mu$, let us compute  
the OPE of the vertex operators ${\cal O}(z)= e^{ip^{(j)}_\mu
X^{(j)}}(z)$
\rschomi,\rseiwit:
\eqn\vertb{ \eqalign{
 e^{i p^{(j)}_\mu X^{(j)}}(\tau)\,   e^{i q^{(j)}_\mu X^{(j)}}
(\tau') &=
           (\tau-\tau')^{{2\alpha' p^{(j)}_\mu q^{(j)}_\mu
            \over 1+
           (F^{(j)}_\mu)^2} }\cr
            &{\rm exp}\left(-i \pi \alpha' {F^{(j)}_\mu \over
            1+(F^{(j)}_\mu)^2 } 
           \epsilon_{kl} p^{(j)}_{\mu k} q^{(j)}_{\mu l}\right) \ 
             e^{i (p^{(j)}_\mu +q^{(j)}_\mu ) X^{(j)}}(\tau')+\ldots .}}
For generic, internal momenta $p^{(j)}_\mu$, 
$q^{(j)}_\mu$ in eq.\kkmom,
the phase factor is indeed nontrivial,
and hence the
torus is noncommutative.  The commutator of the coordinates of the
open string endpoints on the $D9_\mu$-brane is given by
\rchen,\rchu
\eqn\comme{  [ X^{(j)}_1(\tau), X^{(j)}_2(\tau) ]
=-{2\pi i \alpha'  F^{(j)}_\mu \over 
      1+(F^{(j)}_\mu)^2}\, ,}
in agreement with eqs.\commutator,\deforpar.

\subsec{D$(9-d)$-branes at angles}

Instead of working with D9-branes 
with  various magnetic fluxes, we will now use 
the T-dual description in terms of D-branes at angles
\rbgka, which  allows to
present a more intuitive picture of the open string 
sector
involved in such models. 
Applying a T-duality in all $X^{(j)}_2$ 
directions 
\eqn\tdual{  R^{(j)}_2 \rightarrow R^{(j)'}_2 = 1/ R^{(j)}_2 ,}
leads to boundary conditions
for D$(9-d)$-branes intersecting 
at angles, where the angle of the D$(9-d)$-brane relative to
the $X^{(j)}_1$ axes is given by 
\eqn\angles{
\tan\phi^{(j)}=F^{(j)}.} 
(In the following we will omit the prime on
the dual radii.)
This T-duality also maps $\Omega$ onto $\Omega{\cal R}$, 
where ${\cal R}$ acts as complex conjugation on all the $d$  
complex coordinates along the $T^2_{(j)}$ tori. 
Thus, instead of \oria\ we are considering the orientifold
\eqn\orib{   {{\rm Type\, II}\ {\rm on}\ T^{2d} \over \Omega{\cal R} }. } 
For $d$ even we have to take type IIB
and for $d$ odd type IIA. Note that after performing this T-duality
transformation the internal coordinates  are completely commutative.

Let $j\in\{1,\ldots,d\}$ again 
label the $d$ different two-dimensional tori
and $\mu\in\{1,\ldots,K\}$ the different kinds of 
D$(9-d)$-branes, which are distinguished by different angles 
on at least one torus. Moreover,
we are only considering branes which do not densely cover any of the
two-dimensional tori. Thus, 
the position of a D$(9-d)$-brane is described 
by two sets of integers $(n^{(j)}_\mu,m^{(j)}_\mu)$, 
labelling how often
the D-branes are wound around the two fundamental cycles of each $T^2_{(j)}$. 
The 
angles of such a brane with the axes $X_1^{(j)}$ are given by
\eqn\angle{  \tan\phi^{(j)}_\mu={m^{(j)}_\mu R^{(j)}_2 \over 
n^{(j)}_\mu R^{(j)}_1}.}
All these conventions are shown in figure 1. 
\fig{}{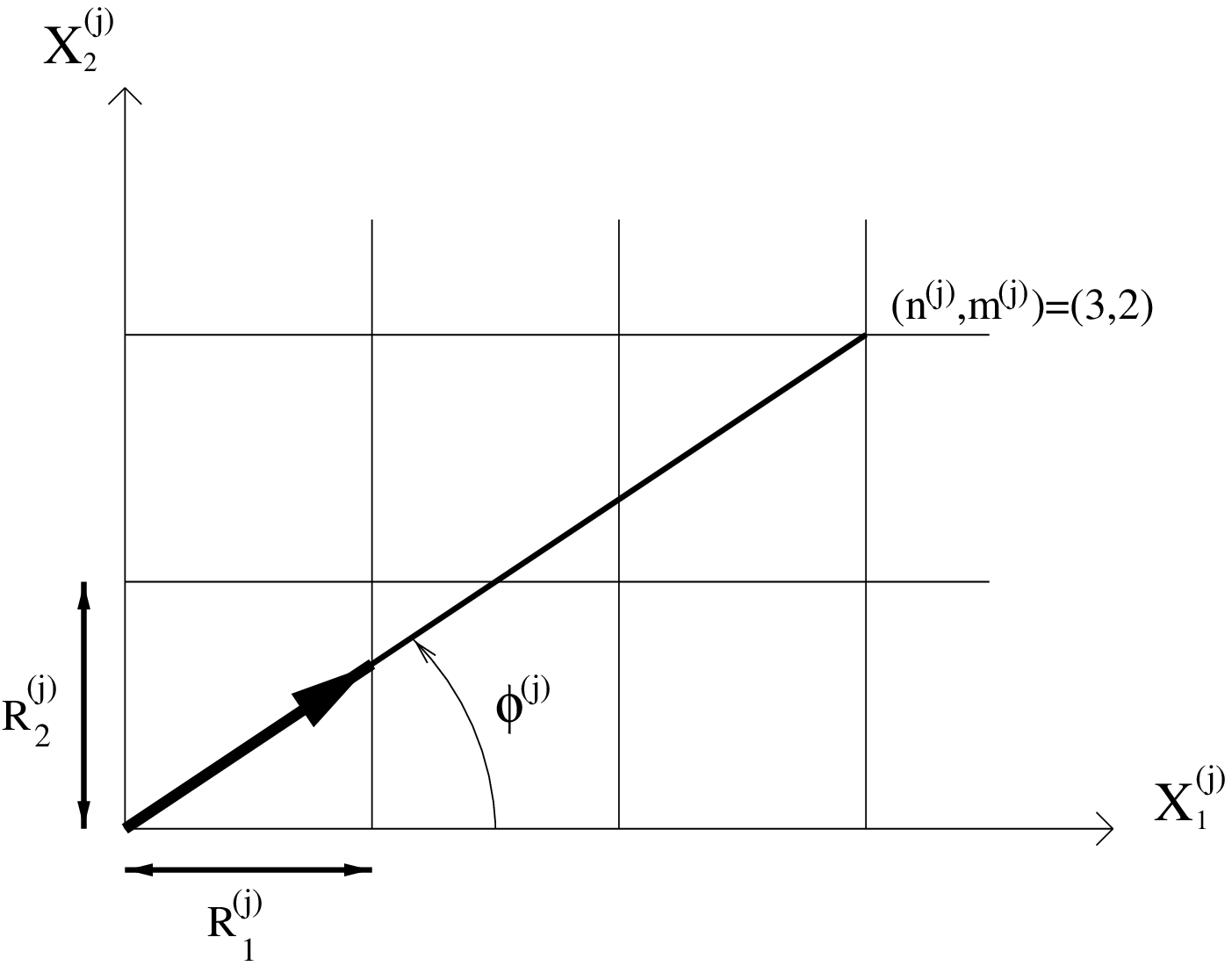}{8truecm}

Since $\Omega{\cal R}$ reflects the D-branes at the axis $X^{(j)}_1$,
for each brane labelled by
$(n^{(j)}_\mu,m^{(j)}_\mu)$ we must also introduce the mirror brane
with $(n^{(j)}_{\mu'},m^{(j)}_{\mu'})=(n^{(j)}_\mu,-m^{(j)}_\mu)$. 
The values $m^{(j)}_\mu =0 \not= n^{(j)}_\mu$ 
and $n^{(j)}_\mu =0 \not= m^{(j)}_\mu$ correspond to branes located along
one of the axis. The horizontal D-branes  translate via T-duality 
into D9-branes with vanishing 
flux and the vertical ones into branes of lower dimension with pure
Dirichlet boundary conditions.

The questions we are going to deal with in the following 
are: Is it possible to cancel 
all or at least the RR tadpoles originating from the Klein bottle amplitude 
by D9-branes with non-vanishing magnetic fluxes $F^{(j)}$,
or equivalently by D$(9-d)$-branes at nontrivial angles $\phi^{(j)}$?  
Taken that supersymmetry is broken generically by such a background, are 
there configuration which still preserve some amount of supersymmetry? 
This would provide a string scenario with partial supersymmetry 
breaking. 
Finally, what are the phenomenological properties of such compactifications?
Concerning the first question we find a positive answer in the sense that 
the RR tadpole can 
be cancelled, while supersymmetry is always broken entirely, albeit sometimes
in  a rather subtle manner. This comes along 
with a non-vanishing NSNS tadpole 
and the presence of tachyons. Interestingly, these models generically contain
chiral fermions motivating us to study how far one can get 
in deriving the standard model in this setting.  
However, later we will mention an 
obstacle to construct phenomenologically 
realistic models in this simple approach.  

Technically we first have to compute all contributions 
to the massless RR tadpole. The cancellation conditions will then imply 
relations for the number of D9-branes and their respective background fluxes. 
This computation will be performed in the T-dual picture, where 
D9-branes with background fields are mapped to D$(9-d)$-branes, and the 
background fields translate into relative angles. This picture 
allows to visualize the D-branes  easily and gives a much better 
intuition than dealing with sets of D9-branes, all filling the same space 
but differing by background fields.

\subsec{Klein bottle amplitude}

The loop channel Klein bottle amplitude for \orib\ can be computed
straightforwardly
\eqn\klein{ {\cal K}=2^{(5-d)} c\, (1-1) \int_0^\infty {dt\over t^{(6-d)}} 
         {1\over 4}
          {\th{0}{1/2}^4 \over \eta^{12} } \prod_{j=1}^d \left[
         \sum_{r,s\in\ZZ} e^{-\pi t \left( {r^2/ \kl{R^{(j)}_1}^2 } +
              s^2 \kl{R^{(j)}_2}^2 \right) } \right] ,}
with $c= {V_{10-2d}}/\left( 8\pi^2 \alpha' \right)^{5-d}$. 
Transforming \klein\ into tree channel, one obtains the
following massless RR tadpole
\eqn\tadkl{  \int_0^\infty  dl\, 2^{(13-d)}  \prod_{j=1}^d \left(
             {R^{(j)}_1\over R^{(j)}_2} \right) .}
The tree channel Klein bottle amplitude allows to determine the 
normalization of the corresponding crosscap states
\eqn\crosscap{ |C\rangle={ 2^{({d/2}-4)} }  
        \left(\prod_{j=1}^d \sqrt{{R^{(j)}_1\over R^{(j)}_2}}\right) \left(
            |C_{\rm NS}\rangle+ |C_{\rm R}\rangle\right) .}

\subsec{Annulus amplitude}

Next we calculate all contributions of open strings stretching between 
the various D$(9-d)$-branes, generically located at nontrivial relative 
angles. We will both  include the case, where the relative angle 
is vanishing, i.e. the background gauge flux is equal on both branes, 
and the case, where the angle is $\pi/2$ and 
the field gets infinitely large on, say, $p$ of the tori. 
 
We start with the contributions of strings with both ends on the same 
brane. 
The T-dual of the  
Kaluza-Klein and winding spectrum in eq.\specb\
reads 
\eqn\kwind{ M_\mu^2=\sum_{j=1}^d \left( \left( {r^{(j)}_\mu 
\over V^{(j)}_\mu} \right)^2 
        + \kl{s^{(j)}_\mu }^2 \left( {R^{(j)}_1 R^{(j)}_2  
\over V^{(j)}_\mu} \right)^2 \right)} 
with 
\eqn\lenght{   V^{(j)}_\mu=\sqrt{ \kl{R^{(j)}_1 n^{(j)}_\mu}^2 
+\kl{R^{(j)}_2 m^{(j)}_\mu}^2 }}
denoting the volume of the brane on $T^2_{(j)}$.
It is now straightforward to compute the loop channel annulus amplitude
for open strings starting and ending on the same brane 
and transform it to the tree channel
\eqn\anntree{ \widetilde{A}_{\mu\mu}=c\, N_\mu^2 (1-1) \int_0^\infty dl\, 
      {1\over 2^{(d+1)}} \prod_{j=1}^d {\kl{V^{(j)}_\mu}^2 
\over R^{(j)}_1 R^{(j)}_2}\ 
            {\th{1/2}{0}^4\over \eta^{12}}
        \sum_{r,s} e^{-\pi l \widetilde{M}^2_\mu} }
with 
\eqn\mtild{ \widetilde{M}_\mu^2=\sum_{j=1}^d 
      \left( {\kl{r^{(j)}_\mu }^2 \kl{V^{(j)}_\mu}^2}
        + \kl{s^{(j)}_\mu }^2  \left( {V^{(j)}_\mu \over  R^{(j)}_1 R^{(j)}_2 }
     \right)^2  \right).}
$N_\mu$ counts the numbers of different kinds of branes.      
Using \anntree\ one can determine the normalization of the boundary
state, which has the schematic form 
\eqn\bound{ |D_\mu\rangle={2^{-({d/2}+1)}}  
        \left(\prod_{j=1}^d {\kl{V^{(j)}_\mu}^2
              \over R^{(j)}_1 R^{(j)}_2}\right)\left(
            |D_{\mu,{\rm NS}}\rangle+ |D_{\mu,{\rm R}}\rangle\right) .}
Reflecting the brane on a single $T^2_{(j)}$ by a $\pi$ rotation onto itself 
corresponds to $(n^{(j)}_\mu,m^{(j)}_\mu)\to (-n^{(j)}_\mu,-m^{(j)}_\mu)$ 
and, as can be determined in the boundary state approach, 
changes the sign of the RR charge, thus exchanging branes and anti-branes. 
Using the  boundary state \bound\ we can compute the tree channel
annulus amplitude for an open string stretched between two different
D-branes                                
\eqn\anndiv{\eqalign{ \widetilde{A}_{\mu\nu}&=\int_0^\infty  dl\,
      \langle D_\mu| e^{-lH_{cl}}| D_\nu\rangle= \cr
&{1\over 2}c\, N_\mu N_\nu  I_{\mu\nu} \int_0^\infty dl 
        \, (-1)^d \sum_{\alpha,\beta\in\{0,1/2\}}  (-1)^{2(\alpha+\beta)}  
         { \th{\alpha}{\beta}^{4-d}\,   \prod_{j=1}^{d}
               \th{\alpha}{(\phi^{(j)}_\nu-\phi^{(j)}_\mu)/ \pi+\beta}\over
                 \eta^{12-3d}\,  \prod_{j=1}^{d}
               \th{1/2}{(\phi^{(j)}_\nu-\phi^{(j)}_\mu)/\pi+1/2} },\cr }}
where the coefficient
\eqn\coeff{ I_{\mu\nu}=\prod_{j=1}^d
           \left( n^{(j)}_\mu m^{(j)}_\nu- m^{(j)}_\mu n^{(j)}_\nu\right) }
is the (oriented) intersection number of the two branes.
It gives rise to an extra multiplicity in the annulus loop channel, which
we have to take into account, when we compute the massless spectrum. 
In order to properly include the case where some 
$\phi^{(j)}_\mu=\phi^{(j)}_\nu$, 
one needs to employ the relation 
\eqn\limit{ \lim\limits_{\psi \rightarrow 0}
             {2 \sin (\pi \psi)  
             \over \th{1/2}{1/2+\psi}} = -{1 \over \eta^3}}
and include a sum over KK momenta and windings as in \anntree.
The contribution to the massless RR tadpole due to \anntree\ and \anndiv\ is
\eqn\tadann{ \int_0^\infty dl\, N_\mu N_\nu\, 2^{(3-d)} 
           \,\prod_{j=1}^d
                 \left( { \kl{R^{(j)}_1}^2 n^{(j)}_\mu n^{(j)}_\nu + 
                       \kl{R^{(j)}_2}^2 m^{(j)}_\mu m^{(j)}_\nu 
                       \over R^{(j)}_1 R^{(j)}_2 }\right) .}
The loop channel annulus can be obtained by a modular transformation
\eqn\annloop{\eqalign{ {A}_{\mu\nu}=c\ N_\mu N_\nu\, I_{\mu\nu}
      \int_0^\infty  {dt\over t^{(6-d)}}\, {1\over 4} \, 
      \sum_{\alpha,\beta\in\{0,1/2\}} &(-1)^{2(\alpha+\beta)}
            e^{2 i \alpha\sum_j (\phi^{(j)}_\nu-\phi^{(j)}_\mu)} 
            e^{i\pi d/2} \cr
               & \times { \th{-\beta}{\alpha}^{4-d}\,   \prod_{j=1}^{d}
               \th{-(\phi^{(j)}_\nu-\phi^{(j)}_\mu)/ \pi-\beta}{\alpha}\over
                 \eta^{12-3d}\,  \prod_{j=1}^{d}
               \th{-(\phi^{(j)}_\nu-\phi^{(j)}_\mu)/\pi-1/2}{1/2} }.}} 
Of course, one can alternatively start from the loop channel, putting in the 
intersection numbers as an extra multiplicity by hand.
The loop channel annulus amplitude looks like a twisted open string sector
and  considering for instance the NS sector, $\alpha=\beta=0$, of \annloop\
one can expand the $\vartheta$-functions in \annloop\ as
\eqn\thfunc{ {\th{0}{0}^{4-d}\over \eta^{12-d}}\, \prod_{j=1}^d\left[
 \left( \sum_{l^{(j)}\ge 1} q^{\epsilon^{(j)} l^{(j)} } \right)
      {\prod_{n^{(j)} \ge 1}(1+q^{n^{(j)}-\epsilon^{(j)} -{1\over 2}})
                            (1+q^{n^{(j)}+\epsilon^{(j)} -{1\over 2}}) \over 
        \prod_{n^{(j)} \ge 1}  (1-q^{n^{(j)}-\epsilon^{(j)}})
              (1-q^{n^{(j)}+\epsilon^{(j)}}) } \right] ,}
with $\epsilon^{(j)}=(\phi^{(j)}_\mu -\phi^{(j)}_\nu)/\pi$.
The non-negative integers $l^{(j)}$ correspond to the  Landau-levels
in \rbac.

\subsec{M\"obius  amplitude}

Computing the overlap between the crosscap state \crosscap\ and a boundary
state \bound\ yields  the contribution of the brane $D(9-p)_\mu$
to the M\"obius amplitude 
\eqn\moediv{\eqalign{ \widetilde{M}_{\mu}=\mp c\, N_\mu\, 2^5\, (-1)^{d}\, 
               & \int_0^\infty dl\,
                  {\prod_{j=1}^d  m^{(j)}_\mu}  \cr
               & \times \sum_{\alpha,\beta\in\{0,1/2\}}  (-1)^{2(\alpha+\beta)}                    { \th{\alpha}{\beta}^{4-d}\,   \prod_{j=1}^{d}
               \th{\alpha}{\phi^{(j)}_\mu/ \pi+\beta}\over
                 \eta^{12-3d}\,  \prod_{j=1}^{d}
               \th{1/2}{\phi^{(j)}_\mu/\pi+1/2} },\cr }}
with argument $q=-{\rm exp}(-4\pi l)$.
Therefore the contribution to the RR tadpole is
\eqn\tadmoe{ \mp \int_0^\infty dl\, N_\mu\,  2^{(9-d)} 
           \,\prod_{j=1}^d \left( {R^{(j)}_1\over R^{(j)}_2} n^{(j)}_\mu 
           \right).}
The overall sign in \moediv\ and \tadmoe\ is fixed by the tadpole
cancellation condition.
In the loop channel the contribution of the M\"obius strip results from 
strings starting on one brane and ending on its mirror partner. 
The extra multiplicity given by the numbers $m^{(j)}_\mu$ of intersection 
points invariant under ${\cal R}$ needs to be regarded as before.  
Now we have all the ingredients to study the relations which derive from 
the cancellation of massless RR tadpoles.

\newsec{Compactifications to six dimensions}

We are compactifying type I strings on a four-dimensional torus and
cancel the tadpoles by introducing stacks 
of D9-branes with magnetic fluxes. 
The T-dual arrangement of D7-branes at angles 
looks like the situation depicted
in figure 2, where we have drawn only two types of D7-branes labelled by 
$\mu$ and $\nu$ and their mirror partners $\mu'$ and $\nu'$, 
the angles being chosen arbitrary.
\fig{}{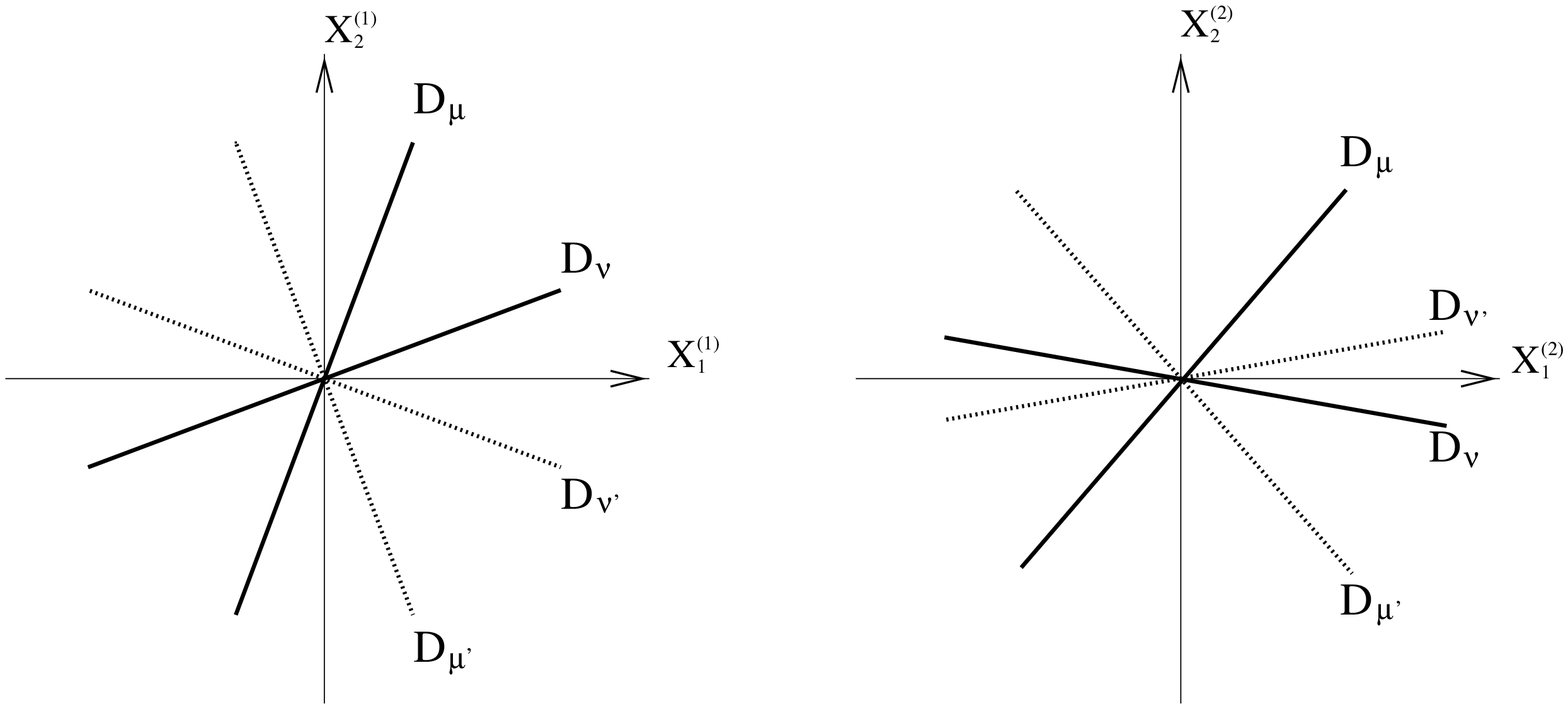}{16truecm}

\subsec{Six-dimensional models}

The complete annulus amplitude is a sum over all open strings stretched 
between the various D7-branes
\eqn\anncom{\eqalign{ \widetilde{A}_{tot}=&\sum_{\mu=1}^K 
        \left( \widetilde{A}_{\mu\mu} +
         \widetilde{A}_{\mu'\mu'}+  \widetilde{A}_{\mu\mu'} +
          \widetilde{A}_{\mu'\mu}\right) + \cr
    &\sum_{\mu<\nu}  \left( \widetilde{A}_{\mu\nu} +\widetilde{A}_{\nu\mu} + 
         \widetilde{A}_{\mu'\nu'}+ \widetilde{A}_{\nu'\mu'} + 
    \widetilde{A}_{\mu\nu'} +\widetilde{A}_{\nu\mu'}+
          \widetilde{A}_{\mu'\nu}+\widetilde{A}_{\nu'\mu} \right) .\cr}}  

Using \anntree\ and \tadann\ and adding up all these various contributions
yields the  following two RR tadpoles 
\eqn\tadannsix{\eqalign{ &\int_{0}^\infty dl\, 8\, 
       \left( {R^{(1)}_1 R^{(2)}_1\over R^{(1)}_2 R^{(2)}_2} \right) \left(
            \sum_{\mu=1}^K N_\mu\,  n^{(1)}_\mu n^{(2)}_\mu
              \right)^2 + 
       \int_{0}^\infty dl\, 8\, 
      \left(  {R^{(1)}_2 R^{(2)}_2\over R^{(1)}_1 R^{(2)}_1}  \right) \left(
            \sum_{\mu=1}^K N_\mu\,  m^{(1)}_\mu m^{(2)}_\mu
              \right)^2 .\cr }}
For the total M\"obius amplitude we obtain the RR tadpole
\eqn\tadmoeb{ \mp \int_0^\infty dl\, 2^8\, {R^{(1)}_1 R^{(2)}_1\over R^{(1)}_2 R^{(2)}_2}\,
         \sum_{\mu=1}^K N_\mu\,   n^{(1)}_\mu n^{(2)}_\mu .}
Note, the two special cases of $N_9$ horizontal and  $N_5$ vertical
D7-branes are contained in \tadannsix\ and \tadmoeb\ by  setting $N_\mu=N_9/2$
respectively  $N_\mu=N_5/2$.
Choosing the minus sign in \tadmoeb\ we get the two RR tadpole 
cancellation conditions
\eqn\tadcancel{\eqalign{ &{R^{(1)}_1 R^{(2)}_1\over R^{(1)}_2 R^{(2)}_2} 
                :\quad\quad \quad
            \sum_{\mu=1}^K N_\mu\, n^{(1)}_\mu n^{(2)}_\mu =16,  \cr
       &  {R^{(1)}_2 R^{(2)}_2\over R^{(1)}_1 R^{(2)}_1} : \quad\quad\quad
            \sum_{\mu=1}^K N_\mu\,  m^{(1)}_\mu m^{(2)}_\mu =0 .\cr }}
As one might have expected, 
pure D9-branes with $m_\mu^{(j)}=0$ only contribute to the tadpole 
proportional to the volume of the torus, while the D5-branes with 
$n^{(j)}_\mu=0$ to the one proportional to the inverse volume.  
Remarkably, by choosing multiple winding numbers, $n^{(j)}_\mu>1$, 
one can  reduce the rank of
the gauge group. As usual in non-supersymmetric models, 
there remains an uncancelled NSNS tadpole, which
needs to be cancelled by a Fischler-Susskind mechanism.

In the next section we shall show that except 
for the trivial 
case, when $m^{(1)}_\mu=m^{(2)}_\mu=0$ for all $\mu$, i.e. vanishing 
gauge flux on all the D9-branes, supersymmetry is
broken and tachyons develop for open strings stretched between different 
branes.  In contrast to the breaking of supersymmetry 
in a brane-antibrane system
these tachyons cannot be removed by turning on Wilson-lines, which is related 
via T-duality to shifting the position of 
the branes by some constant vector. At any 
non trivial angle there always remains an intersection point of two 
D7-branes where the tachyons can localize.    
Also the lowest lying bosonic spectrum depends
on the radii of the torus, which determine the relative angles. 
The zero point energy in the NS sector of a string stretching 
between two different branes is shifted by 
\eqn\vacenergy{ \Delta E_{0,{\rm NS}} = {1 \over 2} 
                \sum_{j=1}^d{ \phi_\mu^{(j)} -\phi_\nu^{(j)} \over \pi}}
using the convention $\phi_\mu^{(j)} -\phi_\nu^{(j)} \in (0,\pi/2]$. 
Even assuming a standard GSO projection, the lightest physical state 
can easily be seen to be tachyonic except for the supersymmetric situation 
with $\phi_\mu^{(1)} -\phi_\nu^{(1)} = \phi_\mu^{(2)} -\phi_\nu^{(2)}$. 
We shall find in the next section that tadpole cancellation prohibits this 
solution, except when all fluxes vanish.   
 
On the contrary, 
the chiral fermionic massless spectrum is independent of the moduli and 
we display it in table 1.
\vskip 0.8cm
\vbox{
\centerline{\vbox{
\hbox{\vbox{\offinterlineskip
\def\tablespace{height2pt&\omit&&\omit&&
 \omit&\cr}
\def\tablerule{\tablespace\noalign{\hrule}\tablespace}

\hrule\halign{&\vrule#&\strut\hskip0.2cm\hfil#\hfill\hskip0.2cm\cr
\tablespace
& spin && rep. && number &\cr
\tablerule
& $(1,2)$ && ${\bf A}_\mu+{\bf \o{A}_\mu}$ &&  $4 m^{(1)}_\mu m^{(2)}_\mu$ 
& \cr
\tablespace
& $(1,2)$ && ${\bf A}_\mu+{\bf \overline{A}_\mu}+
{\bf S}_\mu+{\bf \overline{S}}_\mu$  && 
   $2 m^{(1)}_\mu m^{(2)}_\mu (n^{(1)}_\mu n^{(2)}_\mu -1)$   &\cr
\tablerule
& $(1,2)$ && $({\bf N}_\mu,{\bf N}_\nu)+
   (\o{\bf N}_\mu,\o{\bf N}_\nu)$ && 
 $(n^{(1)}_\mu m^{(1)}_\nu + m^{(1)}_\mu n^{(1)}_\nu)(n^{(2)}_\mu m^{(2)}_\nu 
+ m^{(2)}_\mu n^{(2)}_\nu)$ &\cr
 \tablespace
& $(1,2)$ && $({\bf N}_\mu,\o{\bf N}_\nu)+(\o{\bf N}_\mu,{\bf N}_\nu)$ && 
 $(n^{(1)}_\mu m^{(1)}_\nu - m^{(1)}_\mu n^{(1)}_\nu)(n^{(2)}_\mu m^{(2)}_\nu 
- m^{(2)}_\mu n^{(2)}_\nu)$ &\cr
\tablespace}\hrule}}}}
\centerline{
\hbox{{\bf Table 1:}{\it ~~ Chiral 6D massless open string spectrum.}}}
}
\vskip 0.5cm 
\noindent
(${\bf A}_\mu$ and ${\bf S}_\mu$ denote the antisymmetric resp. symmetric 
tensor representations with respect to $U(N_\mu)$, $SO(N_\mu)$ or $Sp(N_\mu)$.)
Since $\Omega{\cal R}$ exchanges 
a brane with its mirror brane, the Chan-Paton indices of strings ending on 
a stack of branes 
with non-vanishing gauge flux have no $\Omega$ projection and the gauge 
group is $U(N_\mu)$. If $\Omega{\cal R}$ leaves branes invariant, i.e. the 
flux vanishes or is infinite, corresponding to pure D9- or D5-branes, 
the gauge factor is $SO(N_\mu)$ or $Sp(N_\mu)$, respectively.

The degeneracy of states stated in the third column of table 1 
is essentially given by the intersection numbers
of the D7-branes. 
Whenever it is formally negative, one has to pick the $(2,1)$ spinor 
of opposite 
chirality taking into  account the opposite orientation of the branes at 
the intersection. As was pointed out earlier, a change of the orientation 
switches the RR charge in the tree channel translating  into the 
opposite GSO projection in the loop channel.
Therefore  the other chirality survives the GSO projection in the 
R sector. 
If the multiplicity is zero, this does not mean
that there are no massless open string states in this sector, it only
means that the spectrum is not chiral. This happens precisely  
when  two branes lie on top of each other in one of the two $T^2_{(j)}$ tori. 
Then the extra zero modes give rise to an extra spinor state of opposite 
chirality.
The chiral spectrum shown in table 1 does indeed cancel the irreducible
$R^4$ and $F^4$ anomalies. 


We have also considered a $\ZZ_2$ orbifold background, together 
with nonvanishing magnetic flux, which changes the transversality 
condition to 
\eqn\tadcancelorb{
     {R^{(1)}_2 R^{(2)}_2\over R^{(1)}_1 R^{(2)}_1} : \quad\quad\quad
       \sum_{\mu=1}^K N_\mu\,  m^{(1)}_\mu m^{(2)}_\mu =16 }
and leads to a projection $SO(N_\mu),\ Sp(N_\mu) \rightarrow U(N_\mu /2)$ 
on pure D9- and D5-branes but no further changes on D9-branes 
with nonvanishing flux. 
In this background it appears to be possible to construct also 
supersymmetric models \AAD.

\subsec{Supersymmetry Breaking}

One might suspect that there exist nontrivial configurations of D-branes 
cancelling all tadpoles while  preserving  maybe a reduced number of 
supersymmetries.
Now, we would  like to show that no such nontrivial supersymmetric 
configurations exist.
We assume the absence of anti-branes from the beginning as 
their presence breaks supersymmetry anyway. 

Going to the T-dual picture with D-branes at angles, 
we can apply the results of \rangles\ for the 
supersymmetry preserved by two D-branes intersecting at some relative angles 
$\phi^{(1)}$ and $\phi^{(2)}$ on the two tori. Whenever the square of the 
operation $\Theta$ 
which rotates one brane onto the other has eigenvalues 1 when acting 
on the supercharges
\eqn\susycond{ \Theta^2 Q_\alpha =Q_\alpha ,}
some supersymmetry is preserved. Here $Q_\alpha$ denote the sixteen tendimensional 
spinor states with internal quantum numbers $(\pm 1/2,\pm 1/2,...)$ and some definite 
chirality. Hence, $\Theta^2$ has eigenvalues 
$\exp (\pm i\phi^{(1)}\pm i\phi^{(2)} )$ being equal to 1 exactly if 
$\phi^{(1)} =\pm \phi^{(2)}\ {\rm mod}\ 2\pi$, which preserves 
the supercharges $\pm (1/2, -1/2,...)$ or 
$\pm (1/2, 1/2,...)$ respectively. An important point 
to notice is, that the angles must  not be measured modulo $\pi$, but instead 
the relative orientations of the respective 
branes need to be regarded and a consistent 
convention in measuring angles only modulo $2\pi$ in $(-\pi,\pi]$ 
must be adopted. For this purpose, on each $T^2_{(j)}$ we
associate the direction of the vector  
$(n^{(j)}_\mu R_1^{(j)},m^{(j)}_\mu R_2^{(j)})$ to the respective 
D7-brane. 
To preserve supersymmetry, one of the two conditions,  
$\phi^{(1)} =\pm \phi^{(2)}\ {\rm mod}\ 2\pi$, must hold in any 
open string sector. Therefore all branes need to be at equal or 
opposite angles on the two tori.   

Now let us first consider brane configurations without D5-branes, i.e. 
without vertical D7-branes, and concentrate on the second condition in 
\tadcancel, which we call the transversality condition. 
We are free to choose 
$m^{(1)}_\mu\geq 0$.  In order to have any chance to satisfy the transversality
condition, there is at least one pair 
of branes, D7$_1$ and D7$_2$, with  $m^{(2)}_1\geq 0$ and $m^{(2)}_2\leq 0$. 
One can easily convince oneself, that whenever 
D7$_1$ and D7$_2$ have relative angles $\phi^{(1)} =\pm \phi^{(2)}$ 
this cannot be the true for either D7$_1$ and D7$_{2'}$ or D7$_{1'}$ and D7$_2$. 

We still need to include vertical D7-branes to complete the proof. 
They have a positive contribution to the transversality condition in 
\tadcancel. In order to get a net negative contribution from D7-branes at 
angles in $(0,\pi/2)$ relative to the $X^{(j)}_1$ axes, we need 
$\pm m^{(1)}_\mu\geq 0$ and $\mp m^{(2)}_\mu\geq 0$ for some $\mu$. These branes 
have 
$\phi^{(1)} =- \phi^{(2)}$ relative to their mirrors, preserving  
$\pm (1/2, 1/2,...)$, but at best 
may only have $\phi^{(1)} =\pi - \phi^{(2)}$ 
relative to the vertical D7-branes. Thus, such configurations are 
not supersymmetric, either. 
\fig{}{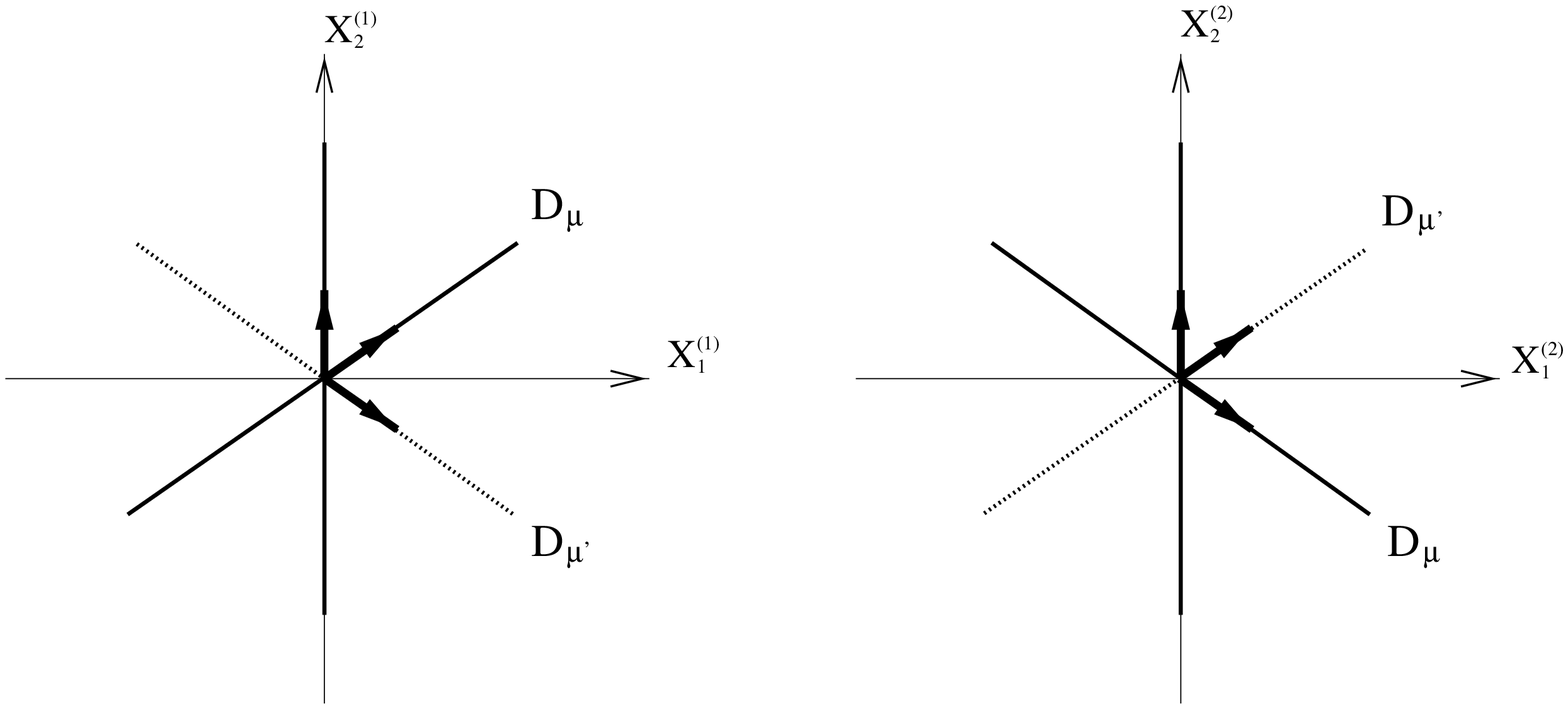}{16truecm}
Such a situation has been illustrated in figure 3. If one flips the orientation of 
the vertical D7-branes on the second torus, the configuration turns 
supersymmetric, as all brane sectors now satisfy 
$\phi^{(1)} = - \phi^{(2)}$. But one finds, that in order to cancel 
the non-abelian anomaly then, one is required to take a different chirality for 
the D5-D9 strings as compared to the rest of the spectrum and therefore the sector 
behaves as in the presence of 
an anti-D5-brane. This refers to the change of the GSO projection 
in the loop channel induced by flipping the orientation.

\newsec{Four dimensional models}

The completely analogous computation as in six dimensions 
can be performed for the compactification of type I strings on a 6-torus 
in the presence of additional gauge fields. 
Now we cancel the tadpoles by D9-branes with magnetic fluxes
on all three 2-tori respectively, in the T-dual picture, 
by D6-branes at angles. 
One obtains four independent tadpole cancellation conditions
\eqn\tadcancelf{\eqalign{ 
&{R^{(1)}_1 R^{(2)}_1 R^{(3)}_1\over R^{(1)}_2 R^{(2)}_2 R^{(3)}_2} :
     \quad\quad \quad
            \sum_{\mu=1}^K N_\mu\, n^{(1)}_\mu n^{(2)}_\mu n^{(3)}_\mu =16  \cr
       &  {R^{(1)}_1 R^{(2)}_2 R^{(3)}_2\over R^{(1)}_2 R^{(2)}_1 R^{(3)}_1} 
             : \quad\quad\quad
            \sum_{\mu=1}^K N_\mu\,  n^{(1)}_\mu m^{(2)}_\mu m^{(3)}_\mu =0 \cr
       &  {R^{(1)}_2 R^{(2)}_1 R^{(3)}_2\over R^{(1)}_1 R^{(2)}_2 R^{(3)}_1} 
             : \quad\quad\quad
            \sum_{\mu=1}^K N_\mu\,  m^{(1)}_\mu n^{(2)}_\mu m^{(3)}_\mu =0 \cr
       &  {R^{(1)}_2 R^{(2)}_2 R^{(3)}_1\over R^{(1)}_1 R^{(2)}_1 R^{(3)}_2} 
             : \quad\quad\quad
            \sum_{\mu=1}^K N_\mu\,  m^{(1)}_\mu m^{(2)}_\mu n^{(3)}_\mu=0. \cr }}
(For convenience they are  given in the  picture with D6-branes at
angles.) 
Again the gauge group contains a $U(N_\mu)$ factor for each stack of 
D9-branes with non-vanishing flux, an $SO(N_\mu)$ gauge factor for 
a stack with vanishing flux and an $Sp(N_\mu)$ factor for a stack of 
D5-branes. 
The general spectrum of chiral fermions with respect to
the gauge group factors is presented in table 2.
\vskip 0.8cm
\centerline{\vbox{
\hbox{\vbox{\offinterlineskip
\def\tablespace{height2pt&\omit&&
 \omit&\cr}
\def\tablerule{\tablespace\noalign{\hrule}\tablespace}

\hrule\halign{&\vrule#&\strut\hskip0.2cm\hfil#\hfill\hskip0.2cm\cr
\tablespace
& rep. && number &\cr
\tablerule
& $({\bf A}_\mu)_L$ &&  $8 m^{(1)}_\mu m^{(2)}_\mu m^{(3)}_\mu$ & \cr
\tablespace
& $({\bf A}_\mu)_L+({\bf S}_\mu)_L$  && 
   $4 m^{(1)}_\mu m^{(2)}_\mu m^{(3)}_\mu 
   (n^{(1)}_\mu n^{(2)}_\mu n^{(3)}_\mu -1)$&\cr
\tablerule
& $({\bf N}_\mu,{\bf N}_\nu)_L$ && 
 $(n^{(1)}_\mu m^{(1)}_\nu + m^{(1)}_\mu n^{(1)}_\nu)(n^{(2)}_\mu 
     m^{(2)}_\nu + m^{(2)}_\mu n^{(2)}_\nu)
 (n^{(3)}_\mu m^{(3)}_\nu + m^{(3)}_\mu n^{(3)}_\nu)$ &\cr
 \tablespace
& $(\o{\bf N}_\mu,{\bf N}_\nu)_L$ && 
 $(n^{(1)}_\mu m^{(1)}_\nu - m^{(1)}_\mu n^{(1)}_\nu)(n^{(2)}_\mu 
   m^{(2)}_\nu - m^{(2)}_\mu n^{(2)}_\nu)
  (n^{(3)}_\mu m^{(3)}_\nu - m^{(3)}_\mu n^{(3)}_\nu)$ &\cr
\tablespace}\hrule}}}}
\centerline{
\hbox{{\bf Table 2:}{\it ~~ Left-handed 4D massless open string spectrum.}}}
\vskip 0.5cm 
\noindent
Whenever the intersection number in the second column is formally negative, 
one again has to take the conjugate representation. The 
spectrum in table 2 is free of non-abelian gauge anomalies. The remarks 
concerning the bosonic NS part of the spectrum made above for six dimensions 
also apply here. Masses depend on the radii and we have not been able 
to produce an otherwise consistent model free of tachyons or even 
preserving supersymmetry.

In the next subsections we discuss some examples and point out some 
phenomenological issues  for these models.

\subsec{A 24 generation $SU(5)$ model}

Having found a way to break supersymmetry, 
to reduce the rank of the gauge group and to produce chiral spectra
in four space-time dimensions, it is tempting to search in a compact
bottom-up approach for brane
configurations producing  massless spectra close to the 
standard model. The tachyons are not that dangerous from the effective
field theory point of view, as they simply may serve as Higgs-bosons 
for spontaneous gauge symmetry breaking, anticipating a mechanism to generate 
a suitable potential keeping their vacuum expectation values finite. 
In \rbac\ a three generation
GUT model was presented, which we shall  revisit 
in the following. 
The gauge group of the model is $G=U(5)\times U(3)\times U(4)\times U(4)$
with maximal rank, so that we have to choose all $n^{(j)}_\mu=1$. 
The following choice of $m^{(j)}_\mu$ then 
satisfies all tadpole cancellation conditions \tadcancelf:
\eqn\mchoice{   m^{(j)}_\mu=\left(\matrix{ 3 & -5 & 1 & -1 \cr
                                       1 &  1 & -1 & -1 \cr
                                       1 &  1 & 1 & 1 \cr} \right) .}
This configuration of D6-branes is displayed in figure 4, where the mirror 
branes have been omitted. 
The chiral part of the fermionic massless spectrum is shown in table 3.  
\fig{}{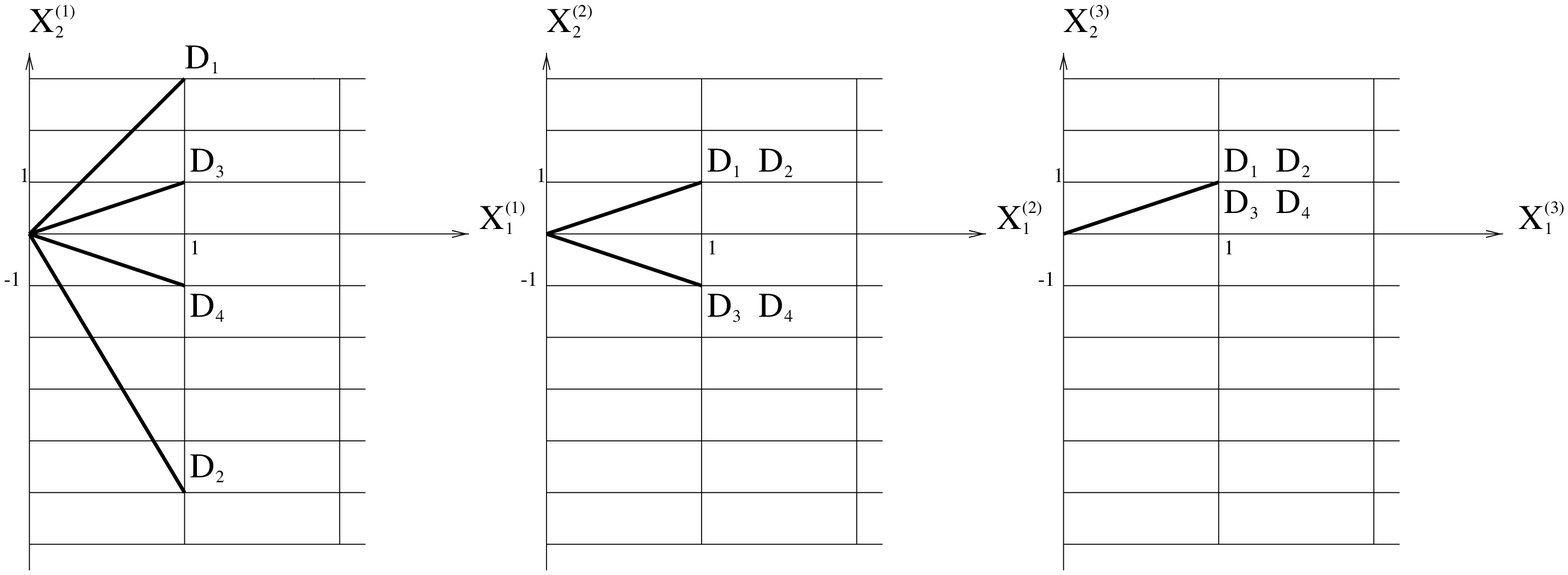}{16truecm}
\vskip 0.8cm
\centerline{\vbox{
\hbox{\vbox{\offinterlineskip
\def\tablespace{height2pt&\omit&&
 \omit&\cr}
\def\tablerule{\tablespace\noalign{\hrule}\tablespace}

\hrule\halign{&\vrule#&\strut\hskip0.2cm\hfil#\hfill\hskip0.2cm\cr
\tablespace
& rep. && number &\cr
\tablerule
& $({\bf 10},{\bf 1},{\bf 1},{\bf 1})$ &&  $24$ & \cr
\tablespace
& $({\bf 1},{\bf 3},{\bf 1},{\bf 1})$ &&  $40$ & \cr
\tablespace
& $(\o{\bf 5},\o{\bf 3},{\bf 1},{\bf 1})$ &&  $8$ & \cr
\tablerule
& $({\bf 1},{\bf 1},\o{\bf 6},{\bf 1})$ &&  $8$ & \cr
\tablespace
& $({\bf 1},{\bf 1},{\bf 1},{\bf 6})$ &&  $8$ & \cr
\tablespace}\hrule}}}}
\centerline{
\hbox{{\bf Table 3:}{\it ~~ Chiral left-handed fermions for the 24 generation 
model.}}}
\vskip 0.5cm 
\noindent
No chiral fermions transform under both the $U(5)\times U(3)$ gauge 
group and the $U(4)\times U(4)$ gauge group, but there will of course also be
non-chiral bifundamentals. If we think of the $SU(5)$ factor as a 
GUT gauge group, then this model has 24 
generations\footnote{$^4$}{In \rbac\ this
model was advocated as a three generation model. We can formally
reproduce the model in \rbac\ by dividing the matrix \mchoice\ by a factor
of two. However, this is inconsistent as it 
would violate the condition that the $m^{(j)}_\mu$'s
have to be integers. Thus, we conclude that in string theory only
the choice \mchoice\ is correct and the model is actually a 24 generation 
model.}. We shall see in the following that it is actually impossible
to get a model with three or any odd number of generations in this framework.

\subsec{A four generation model}

The tadpole cancellation condition 
\eqn\tadcancelfb{ \sum_{\mu=1}^K N_\mu\, n^{(1)}_\mu n^{(2)}_\mu n^{(3)}_\mu 
=16 }
tells us that we can reduce the rank of the gauge group 
right from the beginning by choosing some $n^{(j)}_\mu>1$. Therefore, 
we can envision a model where we start with the gauge group
$U(3)\times U(2)\times U(1)^r$ at the string scale.
In order to have three quark generations in the $({\bf 3},{\bf 2})$ 
representation of $SU(3)\times SU(2)$,
we necessarily need $I_{12}=3$ and $I_{12'}=0$. However, this is not possible,
as $I_{\mu\nu}-I_{\mu\nu'}$ is always an even integer.  The model we found
closest to the 4 generation standard model is presented in the following.
We choose the gauge group $U(3)\times U(2)\times U(1)^2$ and the 
following configuration of four stacks of D-branes: 
\eqn\nchoiceb{   n^{(j)}_\mu=\left(\matrix{ 1 & 1 & 1 & 1 \cr
                                        1 & 1 & 1 & 1 \cr
                                       1 &  1 & 1 & 10 \cr} \right),\quad\quad
                 m^{(j)}_\mu=\left(\matrix{ 0 & 2 & 2 & 0 \cr
                                        0 & 1 & -2 & 0 \cr
                                       1 &  0 & 0 & 1 \cr} \right) .}
The configuration has been illustrated in figure 5.
\fig{}{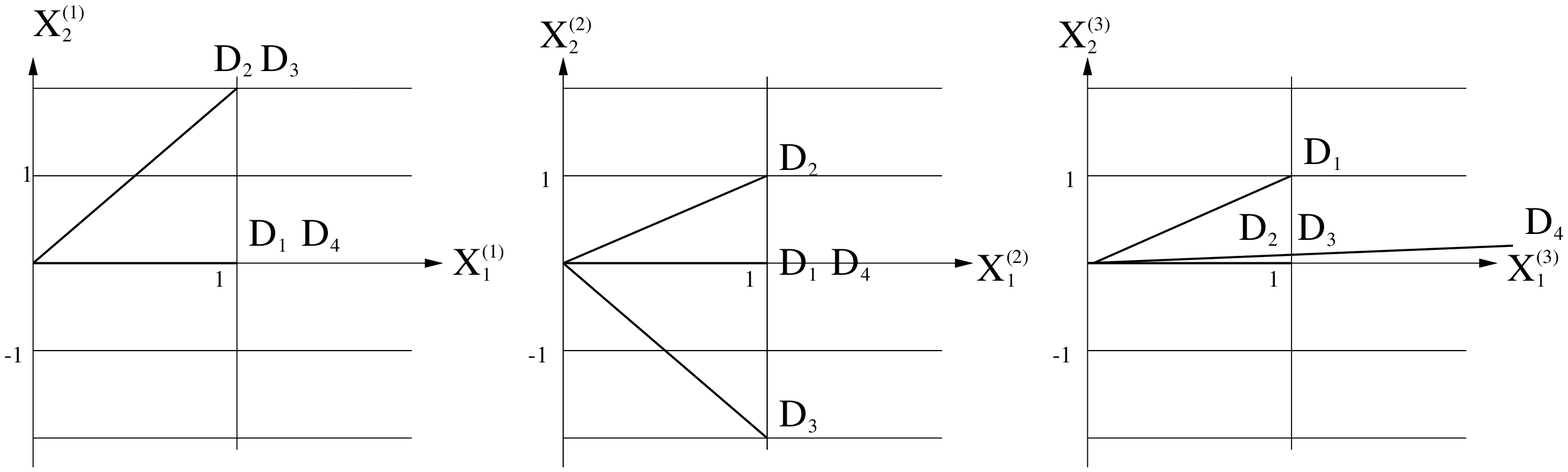}{16truecm}
The resulting chiral massless spectrum is shown in table 4.
\vskip 0.8cm
\vbox{
\centerline{\vbox{
\hbox{\vbox{\offinterlineskip
\def\tablespace{height2pt&\omit&&
 \omit&\cr}
\def\tablerule{\tablespace\noalign{\hrule}\tablespace}

\hrule\halign{&\vrule#&\strut\hskip0.2cm\hfil#\hfill\hskip0.2cm\cr
\tablespace
& $SU(3)\times SU(2)\times U(1)^4$ && number &\cr
\tablerule
& $({\bf 3},{\bf 2})_{(1,1,0,0)}$ &&  $2$ & \cr
\tablespace
& $({\bf 3},{\bf 2})_{(1,-1,0,0)}$ &&  $2$ & \cr
\tablespace
& $(\o{\bf 3},{\bf 1})_{(-1,0,-1,0)}$ &&  $4$ & \cr
\tablespace
& $(\o{\bf 3},{\bf 1})_{(-1,0, 1,0)}$ &&  $4$ & \cr
\tablerule
& $({\bf 1},{\bf 2})_{(0,1,0,1)}$ &&  $2$ & \cr
\tablespace
& $({\bf 1},{\bf 2})_{(0,-1,0,1)}$ &&  $2$ & \cr
\tablespace
& $({\bf 1},{\bf 1})_{(0,0,-1,-1)}$ &&  $4$ & \cr
\tablespace
& $({\bf 1},{\bf 1})_{(0,0, 1,-1)}$ &&  $4$ & \cr
\tablespace}\hrule}}}}
\centerline{
\hbox{{\bf Table 4:}{\it ~~ Chiral left-handed fermions for the 4 generation 
model.}}}
}
\vskip 0.5cm 
\noindent
Computing the mixed $G^2-U(1)$ anomalies one realizes that one of the
abelian gauge factors is anomalous, which needs to be cured by the
Green-Schwarz mechanism. The other three anomaly-free abelian gauge groups
include a suitable hypercharge $U(1)$  
\eqn\hyper{   U(1)_Y={1\over 3} U(1)_1 + U(1)_3 - U(1)_4 ,}
so that the spectrum finally looks like the one in table 5.
\vskip 0.8cm
\centerline{\vbox{
\hbox{\vbox{\offinterlineskip
\def\tablespace{height2pt&\omit&&
 \omit&\cr}
\def\tablerule{\tablespace\noalign{\hrule}\tablespace}

\hrule\halign{&\vrule#&\strut\hskip0.2cm\hfil#\hfill\hskip0.2cm\cr
\tablespace
& $SU(3)\times SU(2)\times U(1)_Y\times U(1)^2$ && number &\cr
\tablerule
& $({\bf 3},{\bf 2})_{({1\over 3},1,0)}$ &&  $2$ & \cr
\tablespace
& $({\bf 3},{\bf 2})_{({1\over 3},-1,0)}$ &&  $2$ & \cr
\tablespace
& $(\o{\bf 3},{\bf 1})_{(-{4\over 3},0,-1)}$ &&  $4$ & \cr
\tablespace
& $(\o{\bf 3},{\bf 1})_{({2\over 3},0, 1)}$ &&  $4$ & \cr
\tablerule
& $({\bf 1},{\bf 2})_{(-1,1,0)}$ &&  $2$ & \cr
\tablespace
& $({\bf 1},{\bf 2})_{(-1,-1,0,)}$ &&  $2$ & \cr
\tablespace
& $({\bf 1},{\bf 1})_{(0,0,-1)}$ &&  $4$ & \cr
\tablespace
& $({\bf 1},{\bf 1})_{(2,0,1)}$ &&  $4$ & \cr
\tablespace}\hrule}}}}
\centerline{
\hbox{{\bf Table 5:}{\it ~~ Chiral left-handed fermions for the 4 generation 
model.}}}
\vskip 0.5cm 
\noindent
We found a semi-realistic, non-supersymmetric, 
four generation standard-model like spectrum 
with two gauged flavour symmetries and right-handed neutrinos. 
In order to determine the Higgs sector, we would have to investigate the bosonic
part of the spectrum. However, this is not universal but depends
on the radii of the six-dimensional torus. We will not elaborate
this further but instead discuss another important issue concerning the 
possible phenomenological relevance of these models. \pano
Since we break supersymmetry already at the string scale $M_s$, in order
to solve the gauge hierarchy problem we must choose 
$M_s$  in the TeV region. Let us employ the T-dual picture 
of D6-branes at angles again to analyze the situation in more detail. 
Using the relations
\eqn\scala{M_{pl}^2 \sim {M_s^8 V_6\over g_s^2}, \quad  
        {1\over (g^{(\mu)}_{\rm YM})^2} \sim {M_s^3 V_\mu\over g_s}, }
where $V_\mu$ denotes the volume of some D6-brane in the internal
directions
\eqn\vol{    V_\mu=\prod_{j=1}^3 V^{(j)}_\mu }
and $g^{(\mu)}_{\rm YM}$ the gauge coupling on this brane. 
They imply
\eqn\rela{  M_s\sim \alpha_{\rm YM}^{(\mu)} M_{pl} 
         {V_\mu\over \sqrt{V_6} } ,}
Therefore, for the TeV scenario to work one needs
\eqn\condi{   {V_\mu\over \sqrt{V_6} }<<1 }
for all D6-branes. However, chirality for the fermionic spectrum of 
an open string stretched between
any two D6-branes implied that the two branes in question do not lie on 
top of each other on any of the three $T^2_{(j)}$ tori. 
In other words the two branes already span the entire torus 
and the condition \condi\ cannot be realized. 

\newsec{Conclusions}

In this paper we have investigated type I string compactifications
on noncommutative tori, which are due to constant magnetic fields along the 
world volumes of the D9-branes being wrapped around the internal space.
The key features of these models in six and four dimensions are the
presence of chiral fermions, low rank gauge groups and broken space-time
supersymmetry. In fact, we found a four-dimensional model with
standard model gauge group $SU(3)\times SU(2)\times U(1)_Y$
(times some Abelian  flavour  gauge groups) with four generations
of standard model fermions and also a four-dimensional 24 generation
$SU(5)$ GUT-like
model.
On the other hand, while displaying promising features like 
getting chiral spectra, these simple models suffer from other 
phenomenological difficulties, like even numbers of 
generations and the problem of splitting the string and the Planck 
scale  sufficiently.  It remains to be seen whether more complicated 
backgrounds with magnetic flux can improve this situation. 

Since the deformation parameters of the noncommutative tori,
we have considered, correspond
to rational (up to some trivial volume dependence) magnetic fluxes,
it was always possible  to choose a T-dual description where
the internal space is commutative, but instead the various D-branes intersect
at particular rational angles. It would be interesting to find also
noncommutative compactifications which do not allow for an equivalent,
T-dual commutative description. It would be also
interesting to discuss the properties of the effective quantum field theories
which originate from noncommutative compactifications along the lines
of ref.\wise.

\vskip 1cm

\centerline{{\bf Acknowledgements}}\pano
The group is supported in part by the EEC contract
ERBFMRXCT96-0045. 
B. K. also wants to thank the Studienstiftung des deutschen Volkes for 
support.

\vskip 1cm

\listrefs

\bye